\documentclass[lettersize,journal]{IEEEtran}
\usepackage{amsmath,amsfonts}
\usepackage{algorithmic}
\usepackage{algorithm}
\usepackage{array}
\usepackage[caption=false,font=normalsize,labelfont=sf,textfont=sf]{subfig}
\usepackage{textcomp}
\usepackage{stfloats}
\usepackage{url}
\usepackage{verbatim}
\usepackage{graphicx}
\usepackage{cite}
\usepackage{xcolor}

\newtheorem{theorem}{Theorem}
\newtheorem{lemma}{Lemma}
\newtheorem{proposition}{Proposition}

\newtheorem{definition}{Definition}

\usepackage{cleveref}

\usepackage{url}

\usepackage{balance}

\begin{document}

\title{When Mobile Crowdsourcing Meets Queueing Systems: {Human-in-the-Loop Learning}}

\author{Hongbo Li,~\IEEEmembership{Member,~IEEE}, Lingjie Duan,~\IEEEmembership{Senior Member,~IEEE}, and Ness B. Shroff,~\IEEEmembership{Fellow,~IEEE}
\thanks{This work has been supported in part by the Army Research Laboratory under Cooperative Agreement Number  W911NF-23-2-0225,  by the U.S. National Science Foundation under the grants: NSF AI Institute (AI-EDGE) 2112471, CNS2312836, CNS-2225561, and CNS-2239677, and by the Office of Naval Research under grant N00014-24-1-2729. The views and conclusions contained in this document are those of the authors and should not be interpreted as representing the official policies, either expressed or implied, of the Army Research Laboratory or the U.S. Government. The U.S. Government is authorized to reproduce and distribute reprints for Government purposes notwithstanding any copyright notation herein.  (Corresponding author: Ness B. Shroff.)}
\thanks{Hongbo Li and Ness B. Shroff are with The Ohio State University, Columbus, OH 43210, USA (emails:li.15242@osu.edu; shroff.11@osu.edu).}
\thanks{Lingjie Duan is with Singapore University of Technology and Design, Singapore 487372 (email: lingjie\_duan@sutd.edu.sg; swing.dlj@gmail.com).}
}

\markboth{IEEE TRANSACTIONS ON NETWORKING}%
{Shell \MakeLowercase{\textit{et al.}}: A Sample Article Using IEEEtran.cls for IEEE Journals}

\IEEEpubid{0000--0000/00\$00.00~\copyright~2021 IEEE}

\maketitle

\begin{abstract}
In today's environment, it is increasingly important for a customer to learn the congestion information of queueing systems (e.g., restaurants, Disneyland attractions, and road networks) before choosing which server to join. {We refer to this paradigm as human-in-the-loop learning (HILL).} Such congestion information becomes outdated over time and emerging crowdsourcing platforms (e.g., Queue-dodging and Waze) are developed for customers to share this information with each other. However, in practice, a selfish customer is unwilling to try other servers with less expected service utility to learn useful information for future customers. 
When mobile crowdsourcing meets queueing systems, it is critical to incentivize HILL among selfish customers to dynamically change their myopic server choices in order to achieve the optimal exploration-exploitation trade-off. 
Our analytical result of infinite price of anarchy (PoA) implies that if customers make myopic choices on which server to select, it can result in an arbitrarily large efficiency loss. This occurs because they overexplore the server even though it may be congested. 
{Specifically, we prove that the lower bound of PoA decreases as the buffer size increases in the single-server case, while the upper bound of PoA decreases as the number of servers increases in the multi-server setting.} As the server choices of customers internally alter the queueing status, we prove that prior informational (non-monetary) mechanisms on exploration-exploitation of exogenous information make the PoA infinite. 
Hence, we propose a dynamic side-payment mechanism, which periodically charges some customers and rewards others to curb their over-exploration over time while keeping ex-post budget balanced. 
Our mechanism balances the congestion and information learning at variable servers such that PoA is reduced to less than $2$. 
In addition to the worst-case PoA analysis, we conduct experiments using real datasets to further verify that our mechanism results in good average performance.
\end{abstract}


\section{Introduction}

In today's queueing systems like restaurants, Disneyland attractions, and road networks, it is crucial for customers to know up-to-date service information for selecting a server. 
This information includes congestion data like server availability \cite{ouyang2012asymptotically} and queue length \cite{hassin2020queue}, which change dynamically based on customers' server choices and service time distribution. 
It is not easy to learn real-time service information, given the tremendous costs of deploying personnel or remote sensors to monitor many servers (\!\!\cite{wang2020efficient,chen2022food}). Recently, crowdsourcing platforms have emerged as a promising way for customers to learn and share congestion information in queueing systems for the public good. 
The Queue-dodging app in London, for example, invites customers to share their estimated waiting time for tables at no-booking restaurants, allowing others to decide whether to join the queue or seek other options \cite{queue-dodging}. 
Disneyland Park apps in Paris and California ask customers to report their waiting times for popular attractions based on their observed queue lengths, providing helpful information for future customers' attraction choices \cite{Disneyland}. Similar platforms for customer-generated information sharing exist for other service applications, such as Pavemint for parking spots \cite{pavemint}, Public Bike System for bike availability and empty dock locations \cite{zhang2018distributed}, Waze for the advisory of less congested paths \cite{waze}, and MyTSA app for estimating waiting time at airport security checkpoints \cite{myTSA}.

Such crowdsourcing platforms expect customers to frequently explore various servers to keep the customer-generated information updated (\!\!\!\cite{wang2020efficient}). To improve learning efficiency, researchers have formulated multi-armed bandit (MAB) problems to determine the optimal policy that balances exploration and exploitation (\!\!\!\cite{slivkins2019introduction}). In MAB problems, an arm (e.g., a server) is selected from multiple arms and an immediate reward is collected to learn about it (e.g., \cite{krishnasamy2021learning,wang2020restless,liu2023nonstationary,li2023congestion,li2020multi,li2025competitive,li2024distributed}). For example, \cite{liu2023nonstationary} proposes a predictive exploration algorithm to decide the current arm choice, enhancing the long-term learning performance for stochastic MAB.  
Similarly, \cite{wang2020restless} studies a restless Markov chain model for each arm and devises an explore-then-commit policy to mitigate performance loss. \cite{krishnasamy2021learning} addresses a simple queueing model with binary service rewards, leveraging MAB techniques to estimate the unknown service rates of individual servers.
However, all of these works assume that customers are altruistic and always follow the platform's socially optimal recommendations upon arrival. This assumption aligns with existing literature on the control of queues (e.g., \cite{stidham1985optimal,legros2018m}).

\IEEEpubidadjcol

In many common queueing systems, customers often make selfish decisions, caring only for their own immediate service utilities upon arrival (\!\!\cite{li2019recommending,chen2022food,gaitonde2021virtues}). 
They are unwilling to try other servers with less expected service utility to learn fresh information for future customers to exploit. 
For example, \cite{li2019recommending} considers a simple road network with one deterministic path (server) and one stochastic path. 
It shows that selfish customers disregard the optimal path recommendations of the social planner, making the system suboptimal without enough exploration of the stochastic path. 
\cite{chen2022food} similarly models the two servers as food delivery (with deterministic utility) and restaurant (with stochastic queue), respectively. 
They find that social welfare will drop if many selfish customers myopically choose delivery services. Hence, it is critical to dynamically change customers' myopic decisions to align with the social optimum. 

When mobile crowdsourcing meets stochastic queueing systems, customers generate both information and congestion for the following users.
Then the key question in our paper is how to analyze and regulate selfish customers' information learning and service progress. There are two technical challenges for us to overcome.
\begin{itemize}
    \item The first is to handle \emph{endogenous information variation in the exploration-exploitation process}. {While many works in the delay announcement literature assume that the service provider actively announces real-time waiting times or queue lengths to customers (\!\!\cite{ibrahim2018sharing}), we focus on information learned and shared by customers themselves.}
    In our problem, the customers' server choices internally alter the dynamic congestion information (e.g., queue length) that is being learned. {Such HILL learning paradigm} is also very different from learning weather or other exogenous information (e.g., \cite{farhadi2022dynamic,mansour2022bayesian}). 
    When positive information learning meets negative congestion, selfish customers may not only under-explore a variable server but also over-explore it over time, adding further difficulty in determining and analyzing the best exploration-exploitation policy. While recent queueing management literature also looks at customer-generated information sharing (e.g., \cite{wang2020efficient,guo2022signaling}), they do not consider any mechanism design to regulate selfish customers.
    \item The other challenge is how to design \emph{a dynamic side-payment mechanism to achieve high efficiency and budget balance}. To remedy the efficiency loss due to selfish customer behaviors, one should design a dynamic incentive mechanism to approach the social optimum as closely as possible. Under endogenous information dynamics, customers may easily reverse-engineer the system states, resulting in the failure of informational mechanisms (e.g., information hiding in Bayesian persuasion \cite{mansour2022bayesian,farhadi2022dynamic}).
    To address this issue, we consider side-payment mechanism design while attaining ex-post budget balance for the crowdsourcing platform, ensuring its long-term sustainability, and ex-ante individual rationality for customers to incentivize their continued participation (\!\!\cite{borgers2015introduction,li2017dynamic}). Since we need to keep the budget no less than zero upon any customer arrival, our mechanism is different from one-shot pricing designs among static customers as in \cite{ferguson2021effectiveness} and \cite{ma2021spatio}.
\end{itemize}

In this work, we will overcome the aforementioned challenges. Our main contributions are summarized next.
\begin{itemize}
    \item \emph{To motivate {human-in-the-loop learning (HILL)} of queueing systems:} To the best of our knowledge, this paper is the first to study how to incentivize HILL among customers to efficiently explore and exploit dynamic queueing systems, where the congestion information varies endogenously with customers’ server choices over time. When mobile crowdsourcing meets queueing systems, we extend traditional one-shot congestion games (e.g., \cite{tavafoghi2017informational,zhang2020traffic,farhadi2022dynamic}) in a fundamental way to create customer's positive learning externalities. We aim to leverage positive information learning to counteract the negative congestion among selfish customers.
    {Our analysis aims to provide insights for the growing class of service systems that rely on decentralized, customer-driven HILL rather than centralized announcements.}
    \item \emph{Threshold-based myopic and socially optimal policies for PoA analysis:} To best adapt to information variation, we formulate the optimization problems as Markov decision processes for both myopic and socially optimal policies. Then we conduct an analytical comparison of these two policies. 
    Our result of infinite price of anarchy (PoA) shows that customers’ myopic server choices can result in an arbitrarily large efficiency loss when compared to the social optimum. In particular, we show that the myopic policy misses both exploitation (when the reported queue length is short) and exploration (when the queue length is long). This finding is distinct from the pure under-exploration result in queueing service literature (e.g., \cite{li2019recommending,chen2022food}).
    \item \emph{Dynamic side-payment mechanism with ex-post budget balance}: 
    To remedy the huge efficiency loss due to myopic behaviors, we first prove that prior informational (non-monetary) mechanisms on exploration-exploitation of exogenous information make PoA infinite, as customers' server choices internally alter the queueing status. 
    Accordingly, we propose a dynamic side-payment mechanism, which periodically charges some customers and rewards others to curb their over-exploration over time while keeping ex-post budget balanced. Our mechanism balances the congestion and information learning at variable servers to reduce the PoA to less than $2$. In addition to the worst-case PoA analysis, we conduct experiments using real datasets to further verify that our mechanism results in good average performance.
\end{itemize}

The rest of the paper is organized as follows. In \Cref{section2}, we formulate the queueing information learning model and customers' service utility model for the system. Then we formulate the optimization problems for both myopic policy and socially optimal policy in \Cref{section3}. After that, in \Cref{section4} we analytically compare the two policies via PoA analysis. Based on the analysis, we propose our dynamic side-payment mechanism in \Cref{section5}. Subsequently, we verify the average-case performance of our side-payment mechanism in \Cref{section5-3}. Finally, \Cref{section6} concludes the paper. For ease of reading, we summarize all the key notations in \Cref{notation_table}.

\section{System Model}\label{section2}

\begin{figure}[t]
    \centering
    \includegraphics[width=0.48\textwidth]{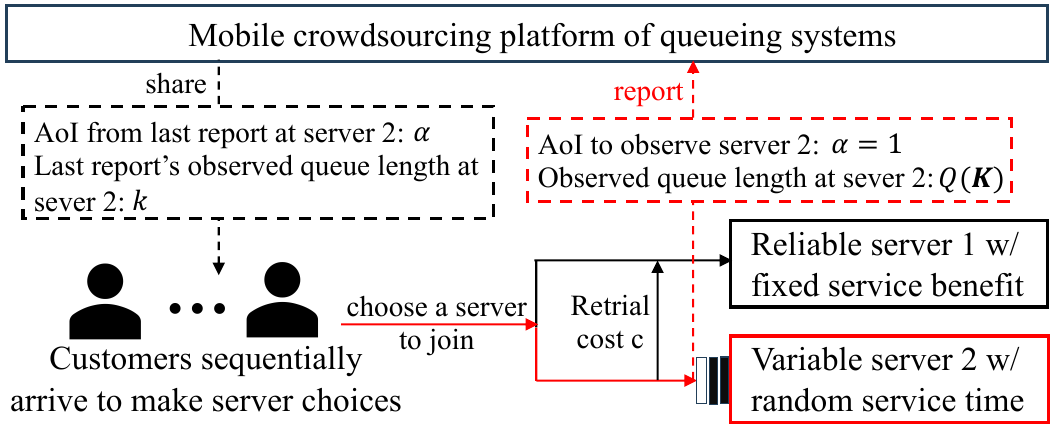}
    \caption{Customers sequentially arrive to request service between a variable server with stochastic service time $F(\mu)$ and a reliable server with fixed service benefit. At the beginning of time $t\in\{1,2,\cdots\}$, a new customer arrival decides the best server (e.g., variable server 2 in this example) to join according to the latest information shared by the crowdsourcing platform. Then it travels to server 2 to learn the actual congestion information and report back to the platform.}
    \label{fig:system_model}
\end{figure}

As illustrated in Figure \ref{fig:system_model}, we consider a crowdsourcing platform of queueing systems lasting for an infinite discrete time horizon. {We discretize time so that, on average, only one customer arrives per slot. This is without loss of generality, as the slot length can be chosen to match the typical customer inter-arrival time in real systems.} At the beginning of each time slot $t\in\{1,2,\cdots\}$, a customer appears to demand food or healthcare service from reliable server~1, which provides fixed service benefits, or a variable server~2, which has random service time/benefit. For example, a customer may choose to prepare its own meal, with a fixed cooking time, or opt for food delivery from a restaurant with random total service times.
For ease of exposition, we first focus on a single variable server here but will extend our analysis to include $N$ variable servers in \Cref{section4-3} and \Cref{section5}. We also study customers' random arrivals via experiments in \Cref{section5-3}.

This two-server model is commonly used in the existing queueing literature (e.g., \cite{lin1984optimal,stidham1985optimal,chen2022food}) and congestion game literature (e.g., \cite{tavafoghi2017informational,li2019recommending,farhadi2022dynamic}).
In this queueing system, the variable server has a limited buffer size $K$, {and we define $F(\mu)$ to be the probability distribution of the random service time for variable server 2, with mean service rate $\mu$.}\footnote{We can also turn to a continuous time model with exponential service time. 
There our analysis still yields an infinite PoA as in \Cref{thm:poa} later, and our mechanism's PoA bound in \Cref{thm:incentive_poa} changes from constant $2$ to~$5$.} 
Once the queue length reaches the buffer size there, it can no longer accept any customer. In this paper, we allow $F(\mu)$ to be any log-concave distribution (e.g., geometric, normal, or exponential distribution) as in the related queueing literature \cite{meisling1958discrete,hassin2016rational}.

\subsection{{Human-in-the-Loop} Information Learning}
As shown in Figure \ref{fig:system_model}, variable server 2 can only serve one customer at a time. When a new customer arrives, it does not know the actual queue length of variable server~2 but only the latest queue length $k$ reported by the prior customer and the age of information (AoI) of the reported queue length $k$, which is denoted by $\alpha$. After selecting the server based on this information, it goes there to learn the actual queue length $\tilde{k}$ and reports back to the platform. Then the platform will share the new reported queue length with the next customer arrival. In the platform, the reported queue length $k$ and AoI $\alpha$ are updated after each customer’s arrival, as described below.
\begin{itemize}
    \item If the newly arrived customer expects that server 2, with its limited buffer $K$, is likely to be congested, it will initially choose server 1. Subsequently, the platform increments the AoI of server 2 from $\alpha$ to $\alpha+1$ for the next customer, without updating the queue length $k$ of server 2.
    \item If the new customer chooses variable server~2, \Cref{fig:system_model} shows that the platform resets $\alpha$ to the minimum~$1$. Further, if server~2 has room, the customer enters and reports its observed customer number $k=\Tilde{k}+1$ to the platform. This number includes itself and $\tilde{k}$ customers in both the buffer and service area of server~2. However, if there is no room, this customer leaves for server 1 and still reports the queue length $k = K + 1$, which includes $K$ customers in the buffer and one customer being serviced in server 2.
Similar to \cite{wang2020efficient}, we normalize the return time to server 1 as zero. 
{In the following equation, we define $Q(k)$ to characterize the queue length dynamics to describe how the queue evolves over time:}
\begin{align}
    Q(k)=\min\left\{\tilde{k}+1, K+1\right\}.\label{Q(k)}
\end{align}
\end{itemize}

\begin{table}[!t]
\renewcommand{\arraystretch}{1.3}
\caption{Key notations and their meanings in the paper}
\label{notation_table}
\centering
\begin{tabular}{|c|m{0.33\textwidth}|}
\hline
\textbf{Notation} & \textbf{Meaning}\\
\hline
\hline
$F(\mu)$ & The distribution of the random service time of server~2 with mean service rate $\mu$.\\
\hline
$G(\lambda)$ & The distribution of customers' random arrival intervals with mean arrival rate $\lambda$.\\
\hline
$K$ & The buffer size of variable server~2.\\
\hline
$\alpha$ & The age of information (AoI) at variable server~2.\\
\hline
$k$ & The last reported queue length at variable server~2.\\
\hline
$Q(k)$ & The dynamic function of $k$.\\
\hline
$r_1$ & The fixed service benefit of reliable server~1.\\
\hline
$r_2$ & The fixed service benefit upon entering the service area of variable server 2.\\
\hline
$h$ & The waiting cost per time slot at server~2.\\
\hline
$c$ & The retrial cost for a customer to switch from server~2 to server~1 {when the queue length at server 2 hits the buffer size}.\\
\hline
$\rho$ & The discount factor.\\
\hline
$R_2(\Tilde{k})$ & A customer's discounted utility of choosing server~2 given actual queue length $\Tilde{k}$ there.\\
\hline
$S(\Tilde{k})$ & The total service time in server~2 given $\Tilde{k}$ customers there. \\
\hline
$g(\alpha,k)$ & The expected utility of choosing variable server~2 given AoI $\alpha$ and last reported queue length $k$ there.\\
\hline
$V(\cdot)$ & The long-term $\rho$-discounted value function.\\
\hline
$A^{(m)}(k)$ & The AoI threshold for exploration under the myopic policy.\\
\hline
$A^*(k)$ & The AoI threshold for exploration under the socially optimal policy.\\
\hline
$q$ & The index of charge-then-reward cycle under our dynamic side-payment mechanism.\\
\hline
\end{tabular}
\end{table}

Based on the above, we formally define the dynamic function of AoI (\!\!\cite{bedewy2019minimizing,li2022online}) per customer arrival below:
\begin{align*}
    \alpha=\begin{cases}
        1, &\text{if the new customer chooses server 2,}\\
        \alpha+1,&\text{otherwise.}
    \end{cases}
\end{align*}
{The pair $(\alpha, k)$ serves as the complete state variable for our Markov decision process (MDP) model. The AoI $\alpha$ increases if no new information is reported, and is reset to $1$ when a new report is generated. The queue length $k$ is updated only when a new customer joins server 2 and reports the latest congestion. This state pair is sufficient to describe the entire decision process, as all future events and rewards depend only on the current $(\alpha, k)$. This structure enables us to model customer decisions, utility, and the evolution of the HILL system within a standard MDP framework.}

Thanks to the HILL information learning provided by the crowdsourcing platform, customers can access the latest information before selecting servers. This marks a departure from the centralized queueing control \cite{legros2018m,hassin2020queue} {and those where a service provider guarantees timely and accurate state announcements \cite{ibrahim2018sharing}}. Next, we introduce the service utility model that informs customers' decision-making.

\subsection{Customers' Service Utility Model}
Similar to the existing works \cite{lippman1975applying,lin1984optimal,stidham1985optimal}, we model that each customer without waiting to enter server~1 receives a fixed service utility/benefit $r_1$. 
Similarly, upon entering the service area of variable server~2 after possible waiting, a customer receives a benefit $r_2$. However, it incurs a waiting cost of $h$ per time slot at server~2. 
Additionally, according to \cite{hassin2020queue}, there is a retrial cost $c$ (e.g., time consumption) for switching from server~2 to server 1 {when the queue length at server 2 hits the buffer size}.
Here both service benefit $r_2$ and waiting cost $h$ are exponentially discounted at a rate $\rho\in(0,1)$ over time. For example, a customer's benefit at server 2 reduces from $r_2$ to $\rho r_2$ after waiting for a time slot. 
{Note that $r_1$ can also be defined as the expected discounted utility, which already accounts for the expected service time at server 1}.

Therefore, let $R_2(\Tilde{k})$ denote a new customer arrival's discounted utility in choosing server~2, which is calculated as follows:
\begin{align}
    R_2(\Tilde{k})=\begin{cases}
        \mathbb{E}_{S(\Tilde{k})}\left[\rho^{S(\Tilde{k})}r_2-\sum_{i=0}^{S(\Tilde{k}+1)-1}\rho^i h\right], \ \text{ if }\Tilde{k}\leq K, \\
        r_1-c, \quad\quad\quad\quad\quad\quad\quad\quad\quad\quad\ \text{ if }\Tilde{k}=K+1,
    \end{cases}\label{R2(ki)}
\end{align}
where $S(\Tilde{k})$ is the total service time given $\Tilde{k}$ customers in server~2 under random service time distribution $F(\mu)$. 
In (\ref{R2(ki)}), if this customer enters server~2 with $\tilde{k}\leq K$, it receives a discounted service benefit $\rho^{S(\Tilde{k})}r_2$ after the service completion of the former $\Tilde{k}$ customer. 
Additionally, it incurs an accumulated waiting cost of $\sum_{i=0}^{S(\Tilde{k}+1)-1}\rho^i h$ while waiting for its own service to be completed. However, if server 2 has no room (i.e., $\tilde{k}=K+1$), it returns to server~1 with retrial cost $c$, leading to utility $r_1-c$. 

However, a new customer arrival is not aware of the exact $\tilde{k}$ for estimating $R_2(\tilde{k})$ in (\ref{R2(ki)}). It can only use the reported queue length $k$ and AoI $\alpha$ of the last report (shared by the crowdsourcing platform) to conclude:
\begin{equation}
        g(\alpha,k)={\mathbb{E}[R_2(\tilde{k})|\alpha,k]}=\sum_{\Tilde{k}=0}^{k} \text{Pr}(\Tilde{k}|\alpha,k) R_2(\Tilde{k}), \label{g(A,k)_multi}
\end{equation}
{where $\text{Pr}(\Tilde{k}|\alpha,k)$ is derived under distribution $F(\mu)$. 
For example, if $F(\mu)$ satisfies the geometric distribution, we solve the binomial formula under geometric distribution $F(\mu)$ to derive
\begin{align}
    \text{Pr}(\Tilde{k}|\alpha,k)=\begin{pmatrix}{\alpha}\\{ k-\Tilde{k}}\end{pmatrix}(1-\mu)^{\alpha-(k-\Tilde{k})}\mu^{k-\Tilde{k}}.\label{probability_k}
\end{align}}
According to our information learning model in (\ref{Q(k)}), if this customer chooses to explore server 2, its shared information changes the expected utility in (\ref{g(A,k)_multi}) to $g(1,Q(k))$ for the subsequent customer. Otherwise, this utility becomes $g(\alpha+1,k)$.

\section{Problem formulation for Myopic and Socially Optimal Policies}\label{section3}
Based on information learning and customers' service utility models, we are ready to formulate dynamic optimization problems for customers' myopic server choices and the social planner's optimal server advisory, respectively. 

\subsection{Problem Formulation for the Myopic Policy}
Before formulating the problem for customers, we first define the myopic policy (e.g., used in \cite{li2019recommending,li2023congestion,tavafoghi2017informational}) below.
\begin{definition}[Myopic Policy \cite{li2019recommending,li2023congestion,tavafoghi2017informational}]\label{def:myopic_policy}
Under the myopic policy, each customer selects the server to maximize its own immediate expected utility.
\end{definition}

Based on \Cref{def:myopic_policy}, a customer upon arrival will follow its selfish interest to make a myopic decision.
{In this case, customers only care about their own current expected utility, truthfully reporting the queue length information is incentive-compatible for it (\!\!\cite{li2019recommending,borgers2015introduction}).}

Define $V^{(m)}(\alpha,k)$ as the long-term $\rho$-discounted value function to include the total utility of all the customers under the myopic policy. 
For the myopic policy, a selfish customer myopically compares $r_1$ of server 1 and $g(\alpha,k)$ in (\ref{g(A,k)_multi}) of server~2 to decide which server to join, without caring about the information learning efficiency and the introduced congestion to future customers. 
Hence, the dynamics of $V^{(m)}(\alpha,k)$ per customer arrival only depend on the two following cases.

If $r_1>g(\alpha,k)$, the current selfish customer will choose reliable server~1 and obtain the service utility $r_1$. Then the AoI $\alpha$ in server 2 increases to $\alpha+1$, and the reported queue length remains at $k$, such that the value function of the next customer is $V^{(m)}(\alpha+1,k)$. 

If $r_1\leq g(\alpha,k)$, the customer will choose server~2 and obtain expected utility $g(\alpha,k)$ in (\ref{g(A,k)_multi}). As the AoI is reset to $1$, the expected value function {for the next customer} in this case is:
    \begin{equation}
    \mathbb{E}\big[V^{(m)}\big(1,Q(k)\big)\big]=\sum_{\Tilde{k}=0}^{K+1}{\text{Pr}}\big(\tilde{k}|\alpha,k\big) V^{(m)}\big(1,Q(k)\big),\label{E_V}
\end{equation}
where queue lengths dynamics $Q(k)$ is defined in (\ref{Q(k)}) and probability ${\text{Pr}}\big(\tilde{k}|\alpha,k\big)$ is derived in (\ref{probability_k}).

Accordingly, given discount factor $\rho\in(0,1)$, we can formulate the following dynamic programming problem for the myopic policy to obtain the dynamics of $V^{(m)}(\alpha,k)$:
\begin{align}\label{Multi_V_myopic}
    V^{(m)}(\alpha,k)=\begin{cases}
    r_1+\rho V^{(m)}( \alpha+1,k),\quad\ \ \text{ if } r_1>g(\alpha,k),\\
        g(\alpha,k)+\rho\mathbb{E}\big[V^{(m)}\big(1,Q(k)\big)\big],\text{ otherwise.} 
        \end{cases}
\end{align}

In (\ref{Multi_V_myopic}), due to the myopic behavior, the current customer is not willing to explore variable server 2 if $r_1>g(\alpha,k)$, which will hinder learning fresh information at server 2 for future customers. On the other hand, if $g(\alpha,k)\geq r_1$, this customer will explore variable server 2 to add congestion to future customers in this server. We will analyze the myopic policy and its PoA in \Cref{section4}.

\subsection{Problem Formulation for Social Optimum}
The objective of the socially optimal policy is to maximize the total discounted utility of all the customers in the long run. Note that this is different from the myopic policy that only cares about an arrival’s immediate service utility. Next, we formulate the social welfare maximization problem for all the customers as a Markov decision process (MDP). 

Let $V^*(\alpha,k)$ be defined as the optimal long-term $\rho$-discounted value function. This function dynamically changes per customer arrival, depending on which current choice (of reliable server~1 or variable server 2) yields a greater long-term value.

\begin{itemize}
    \item If the social planner asks the customer to choose reliable server~1, the customer will receive immediate service utility $r_1$. As it does not learn any information (by increasing $\alpha$ by $1$ and keeping the same report $k$), the optimal value function for the next customer is $V^{*}(\alpha+1,k)$.
    \item If the social planner asks the customer to explore variable server 2, then the platform resets the AoI of this server to $\alpha=1$. Depending on whether server 2 has room to enter or not, the optimal expected value function from the next customer will be $\mathbb{E}_{\tilde{k}}[V^*(1,Q(k))]$, which can be determined similarly as in (\ref{E_V}).
\end{itemize}

Based on these two cases, we formulate the optimal value function to maximize social welfare:
\begin{align}
    V^*(\alpha,k)
        =\max&\big\{r_1+\rho V^*( \alpha+1,k),\notag\\ &g(\alpha,k)+\rho\mathbb{E}_{\tilde{k}}\big[V^*\big(1,Q(k)\big)\big]\big\}.\label{Multi_V*}
\end{align}

By comparing with the myopic policy in (\ref{Multi_V_myopic}), the optimal value function (\ref{Multi_V*}) considers the long-term utility to decide the current server choice.
{In principle, the socially optimal policy can be computed by brute-force enumeration \cite{bellman1966dynamic}. To address the computational burden, various approximation methods from dynamic programming and reinforcement learning can be employed, e.g., approximate dynamic programming \cite{powell2007approximate} and Q-learning \cite{sutton1998reinforcement}. However, since the focus of our work is on analyzing the performance gap between the myopic policy and the socially optimal policy, we do not present the full computational process for obtaining the socially optimal policy in this paper. Instead, we focus on establishing structural results and designing mechanisms to regulate customer behavior.}
In the next section, we provide an analytical comparison between the myopic and socially optimal policies.

\section{Myopic Policy versus Socially Optimal Policy for PoA Analysis}\label{section4}

In this section, we prove that both policies are threshold-based and analyze the monotonicity of their thresholds in the reported queue length and buffer size. We also analyze that the myopic policy misses both exploration and exploitation as compared to the socially optimal policy, leading to an infinite PoA. Finally, we extend to a more general case of multiple variable servers to understand its effect and the resultant PoA.

\subsection{Threshold-based Policy Solutions}
By analyzing problems (\ref{Multi_V_myopic}) and (\ref{Multi_V*}), we first prove both value functions' monotonicity in $\alpha$ and $k$. 
\begin{lemma}\label{lemma:V*_mono}
Both value functions $V^{(m)}(\alpha,k)$ in (\ref{Multi_V_myopic}) and $V^*(\alpha,k)$ in (\ref{Multi_V*}) increase with AoI $\alpha$ and decrease with queue length $k$ of last report at variable server 2.
\end{lemma}

The proof of \Cref{lemma:V*_mono} is given in Appendix~A of the supplemental materials.
{Intuitively, as $\alpha$ increases, more time has passed since the last report at server 2, allowing for more service completions and thus likely reducing the actual queue length compared to the previously reported value. As a result, the expected utility for the next customer increases with $\alpha$ under both policies.
Conversely, as the last reported queue length $k$ increases, the likelihood of congestion and longer waiting times at server 2 rises, which reduces the expected utility and, consequently, the value function.}

Given Lemma \ref{lemma:V*_mono}, we next prove both policies are threshold-based. This guides any customer's server choice after observing the last report's queue length $k$ and AoI $\alpha$ upon arrival. 
\begin{lemma}\label{lemma:thresholds}
Under myopic policy (\ref{Multi_V_myopic}), customer arrivals will keep choosing reliable server 1 instead of variable server~2 until its AoI $\alpha$ from the last visit exceeds the threshold below:  
\begin{equation}
    A^{(m)}(k)= \arg\min_{\alpha}\{\alpha|r_1\leq g(\alpha,k)\}.\label{nm_tau}
\end{equation}
Similarly, the socially optimal policy in (\ref{Multi_V*}) is also of threshold-type, and will recommend customers to explore variable server~2 (instead of staying with server 1) after $\alpha$ exceeds the following threshold:
\begin{align}
        A^{*}(k)=\arg\min_{\alpha}\Big\{\alpha\Big |&r_1+\rho V^*(\alpha+1,k)\leq\label{ns_tau}\\
        &g(\alpha,k)+\rho\mathbb{E}_{\tilde{k}}\Big[V^*\big(1,Q(k)\big)\Big]\Big\}.\notag
\end{align}
Both exploration thresholds $A^{(m)}(k)$ and $A^{*}(k)$ increase with $k$ but decrease with buffer size $K$.
\end{lemma}

The proof of \Cref{lemma:thresholds} is given in Appendix~B of the supplemental material.
According to \Cref{lemma:V*_mono}, customers' expected rewards increase with the AoI $\alpha$. Therefore, after the last visit of variable server 2 that reset $\alpha$ to $1$, new customer arrivals prefer to stay with server 1 and will explore variable server~2 after waiting for certain time thresholds in both policies to achieve larger expected rewards. As reported queue length $k$ becomes larger, customers prefer not to explore this variable server. If buffer size $K$ increases, it has less probability for customers to be blocked by server~2, such that customers become more willing to explore.

\subsection{Policy Comparison and PoA Analysis}
In this subsection, we analytically compare the exploration thresholds $A^{(m)}(k)$ in (\ref{nm_tau}) and $A^{*}(k)$ in (\ref{ns_tau}) as well as their associated social welfare. 
First, we propose the following lemma to compare the two exploration thresholds' monotonicity in $k$.
\begin{lemma}\label{lemma:Mono_Am-A*}
   Given fixed buffer size $K$, the difference between the two thresholds, i.e., $A^{(m)}(k)-A^*(k)$, is non-decreasing in the last report's queue length $k$. 
\end{lemma}

The proof of \Cref{lemma:Mono_Am-A*} is given in Appendix~C of the supplemental material.
Intuitively, as the last report's queue length $k$ increases, selfish customers worrying about congestion are unwilling to explore variable server 2, as compared to the social optimum.

Then we define under/over-exploration for the myopic policy as follows.
\begin{definition}[Under/over-exploration]\label{def:under/over-exploration}
The myopic policy under-explores variable server 2 if $A^{(m)}(k)\geq A^{*}(k)$, and over-explores if $A^{(m)}(k)\leq A^{*}(k)$.
\end{definition}

Based on \Cref{lemma:Mono_Am-A*} and \Cref{def:under/over-exploration}, we next prove that, in comparison to the socially optimal policy, the myopic policy misses both exploration and exploitation of variable server 2.

\begin{proposition}\label{prop:explore_sys2}
There exists a threshold $k_{th}\in\{1,\cdots,K+1\}$ such that, when observing a small $k\leq k_{th}$ from the platform, the myopic policy will over-explore variable server~2 (i.e., $A^{(m)}(k)\leq A^{*}(k)$), but may under-explore (i.e., $A^{(m)}(k)\geq A^{*}(k)$) when observing a large $k> k_{th}$, as compared to the social optimum.
\end{proposition}

The proof of \Cref{prop:explore_sys2} is given in Appendix~D of the supplemental material. 
\Cref{prop:explore_sys2} tells that the myopic policy misses both exploitation and exploration, as reported queue length $k$ changes over time. Given a small $k$ of the last report, myopic customers keep choosing this variable server without considering the congestion caused to future customers. 
However, if $k$ is large, they are unwilling to explore due to low expected utility. In this case, the socially optimal policy still recommends that certain customers explore server 2 to learn fresh information to reduce $k$ there. 
So, the socially optimal policy improves the customers' future utility at the cost of the current utility to these customers.
{Note that actual frequency of over-exploration or under-exploration events in practice will depend on the system’s parameter settings and typical operating regimes. For example, in many practical scenarios, long queue states may be rare, so under-exploration could occur less frequently than over-exploration.}

After comparing the thresholds of the two policies, we are ready to examine how the efficiency of this dynamic system degrades due to the myopic policy.
Formally, according to \cite{koutsoupias1999worst}, we define the Price of Anarchy ($\text{PoA}$) as the maximum ratio between the expected total discounted value function $V^*(\alpha,k)$ in (\ref{Multi_V*}) under the socially optimal policy and value function $V^{(m)}(\alpha,k)$ in (\ref{Multi_V_myopic}) under the myopic policy. That is,
\begin{equation}
    \text{PoA}^{(m)}=\max_{r_1,r_2,\mu,\rho,c,\alpha,k,K}\frac{V^*(\alpha,k)}{V^{(m)}(\alpha,k)},\label{PoA_2}
\end{equation}
which is always larger than $1$. We want to find its maximum to characterize the myopic policy's worst-case efficiency loss by searching through all possible system parameters. 

Next, we will prove that $\text{PoA}^{(m)}$ can be arbitrarily large.
\begin{theorem}\label{thm:poa}
The $\text{PoA}^{(m)}$ defined in (\ref{PoA_2}) satisfies the following lower bound
\begin{align}
    \text{PoA}^{(m)}\geq \frac{1+\rho \frac{g(2,{K+1})}{r_1}}{1+\rho},\label{PoAm}
\end{align}
which approaches infinity if the system parameters satisfy $c\gg r_1,r_2\gg r_1,$ and $\mu=\frac{c}{R_2(K)+c-r_1}$ {in $g(2,K+1)$}.
\end{theorem}

The proof of \Cref{thm:poa} is given in Appendix E. This result of $\text{PoA}^{(m)}\geq \infty$ holds if variable server 2 reaches its capacity ($k=K+1$). 
Intuitively, much larger service benefit $r_2\gg r_1$ (e.g., with $r_1\rightarrow 0$) in server 2 and non-small service rate $\mu=\frac{c}{R_2(K)+c-r_1}$ make variable server 2 more attractive than reliable server 1. 
This leads to maximum over-exploration of variable server 2 (e.g., $g(1,K+1)\geq r_1$) to cause rejection there and trigger a large retrial cost (with $c\gg r_1$) to pay. However, the social optimum will strategically ask some customers to choose reliable server~1 instead of jamming variable server 2. 
Then, for the next customer arrival, their expected utility $g(2,K+1)$ at variable server $2$ will obviously increase.
According to (\ref{R2(ki)}), $g(2,K+1)$ decreases with $K$ given fixed service benefit $r_2$, and thus the lower bound of $\text{PoA}^{(m)}$ in (\ref{PoAm}) also decreases with $K$. 

{Since PoA is defined as the maximum efficiency ratio across all possible system parameters, demonstrating a single case where PoA diverges to infinity is sufficient to establish that $\text{PoA} \geq \infty$, i.e., the system can suffer unbounded efficiency loss under myopic behavior.
Consequently, \Cref{thm:poa} motivates us to design an effective mechanism to regulate the myopic policy.}

\subsection{{PoA Analysis for Multiple Variable Servers}}\label{section4-3}
{In the previous analysis, we focused on the system with a single variable server and demonstrated that myopic customer behavior can lead to an arbitrarily large PoA compared to the social optimum. In this subsection, we extend our analysis to the more general case of $N$ independent variable servers, each with identical expected service rate $\mu$, and investigate how the presence of multiple variable servers impacts system efficiency under the myopic policy. 
Even with heterogeneous servers, the PoA lower bound in \eqref{PoAm} still applies. This can be seen by setting $N = 2$ and keeping the other parameters unchanged as in Theorem 1.}

{Consider a system with $N$ variable servers. Upon arrival, each customer selects their preferred server. If their first choice is blocked, they may switch to the next-best server. Specifically, the myopic policy compares the immediate expected utilities of the $N$ left servers (except for the first choice) to make its second decision. On the other hand, the socially optimal policy compares the caused social welfare. Both policies repeat the above decision-making processes until successfully entering a queue with available buffer.
The reported queue length $k_j$ for each variable server $j \in \{2, \ldots, N+1\}$ is updated in the same manner as described previously.}

{While \Cref{thm:poa} established a lower bound for PoA in the single-server case, it is important to understand whether and how increasing the number of servers can mitigate efficiency loss. To this end, we present the following proposition, which characterizes an upper bound for PoA as a function of $N$.}
\begin{proposition}\label{Prop:PoA_N}
    If there are $N$ variable servers, the myopic policy achieves 
    \begin{align*}
        \text{PoA}^{(m)}_N\leq \left\lceil\frac{K+1}{ \max\{\lfloor N\mu-1\rfloor,0\}}\right\rceil,
    \end{align*}
    where $\lceil (\cdot)\rceil$ is the minimal integer that is larger than $(\cdot)$, and $\lfloor (\cdot)\rfloor$ is the maximal integer that is smaller than $(\cdot)$. Given buffer size $K$, $\text{PoA}^{(m)}_N$ approaches to $1$ as $N\rightarrow\infty$.
\end{proposition}

The proof of \Cref{Prop:PoA_N} is given in Appendix~F of the supplemental material.
{With $N$ variable servers, customers successively explore each server $i \in \{2, ..., N+1\}$, with the age of information (AoI) $a_i = N$ for each server. As a result, the expected queue length at a variable server can decrease by approximately $\max\lfloor \{N\mu - 1\rfloor,0\}$ between explorations under the myopic policy. In contrast, the social optimum can potentially reduce the queue length by up to $K + 1$ per exploration. This difference in possible queue reduction explains the upper bound on PoA.}

{This upper bound shows that the worst-case inefficiency of the myopic policy (i.e., maximum PoA) decreases as the number of variable servers increases. In the limit as $N$ grows large, PoA converges to $1$, indicating that having many independent servers protects the system from efficiency loss due to myopic behavior. 
It is also clear from \cref{Prop:PoA_N} that the worst-case inefficiency occurs when $N = 1$.
The intuition is that when only a single variable server is available, all exploration is concentrated on this one server. This can lead to excessive and redundant exploration by myopic customers, who cannot distribute their exploratory actions across multiple servers.
Therefore, our mechanism design efforts in the following sections focus primarily on this challenging single-server setting, aiming to efficiently reduce the potentially unbounded PoA to a constant.}

\section{Budget-balanced Side-payment Mechanism Design}\label{section5}
As shown in \Cref{thm:poa}, the efficiency loss due to myopic customer behaviors can be arbitrarily large. To address this issue, we aim to design an efficient mechanism to regulate selfish customers and reduce infinite PoA to a finite constant. In this section, we first prove that any informational (non-monetary) mechanism fails to work. Then we propose our new dynamic side-payment mechanism and prove its achieved good performance. For ease of exposition, in this section, we first consider $N=1$ for a single variable server, and will extend our analysis to an arbitrary server number $N$ in \Cref{thm:incentive_poa}.

\subsection{Benchmark Mechanisms Comparison}
We first follow mechanism design literature \cite{borgers2015introduction} to formally define informational mechanisms below.
\begin{definition}[Informational Mechanism \cite{borgers2015introduction}]\label{def:information_mechanism}
An informational mechanism establishes a Bayesian game, wherein the social planner aims to align participants' interests with desired objectives through strategic information provision, which may involve information hiding, information disclosure, or strategic recommendations.
\end{definition}

According to \Cref{def:information_mechanism}, the information-hiding mechanism, for example, is commonly employed in the literature (e.g., \cite{farhadi2022dynamic,li2019recommending}). In this work, the information-hiding mechanism can hide both AoI $\alpha$ and reported queue length $k$ from customers while offering optimal server recommendations.
However, we propose the following lemma to show that any informational mechanism (including information hiding) cannot work in our system with endogenous information dynamics.
\begin{lemma}\label{Lemma:Information}
No informational mechanism can result in a bounded PoA, i.e., there exists no finite value $\alpha$ such that an informational mechanism can guarantee $\text{PoA} < \alpha$.
\end{lemma}

The proof of \Cref{Lemma:Information} is given in Appendix~G of the supplemental material.
To prove this lemma, we still consider the worst-case scenario in \Cref{thm:poa}, where selfish customers maximally over-explore variable server 2. 
{Note that each myopic customer chooses between the variable server and the reliable server by comparing the expected utility $g(\alpha, k)$ with $r_1$. Because $g(\alpha, k)$ is increasing in the age of information $\alpha$ and decreasing in the reported queue length $k$, the value $g(1, K+1)$ serves as a lower bound for $g(\alpha, k)$ over all possible system states. Therefore, if $g(1, K+1) \geq r_1$, it follows that $g(\alpha, k) \geq r_1$ for all $(\alpha, k)$. In this situation, no informational mechanism can alter myopic customer choices, since the variable server always appears at least as attractive as the reliable server.}
In addition to the theoretical analysis in \cref{Lemma:Information}, we later conduct experiments on the information-hiding mechanism to evaluate its poor average-case performance in comparison to the socially optimal policy.

We now design a side-payment (monetary) mechanism to remedy the efficiency loss. The platform may charge extra fees from some customers and reward others over time, for altering their service utility objectives in \Cref{section3}. 
In \cite{borgers2015introduction,li2017dynamic}, it is necessary to ensure ex-post budget balance for the platform, and we follow the same rationale here.
\begin{definition}[Ex-post budget balance \cite{borgers2015introduction,li2017dynamic}]\label{def:expost_budget}
A dynamic mechanism is said to be ex-post budget balanced if the platform's budget is no less than zero upon any customer arrival in the time horizon. 
\end{definition}

We cannot borrow the existing one-shot side-payment or pricing mechanisms with budget balance from the literature (e.g., \cite{ferguson2021effectiveness,ma2021spatio}), as they are not designed for a dynamic system with customers' sequential arrivals and random services over time. 
Alternatively, to keep ex-post budget balanced for the platform, we propose to selectively charge some customers following the socially optimal policy, and then use the budget to reward those customers expecting to deviate from the social optimum. 
Our goal is to change myopic customers' exploration threshold $A^{(m)}(k)$ in (\ref{nm_tau}) to approach the optimal exploration threshold $A^*(k)$ in (\ref{ns_tau}) for actual $\alpha$ and $k$ at any time. 

Additionally, we aim for our mechanism to ensure ex-ante individual rationality for customers to encourage their continued participation \cite{kosenok2008individually,alva2020impossibility}. We formally define ex-ante individual rationality below.
\begin{definition}[Ex-ante individual rationality \cite{kosenok2008individually,alva2020impossibility}]\label{def:exante_IR}
{Let $U_t$ be the expected utility for the $t$-th customer under the mechanism, and $U_t^{\text{out}}$ be the utility from opting out. The mechanism is ex-ante individually rational if
\begin{align}\label{EU_exante}
    \mathbb{E}[U_t]\geq U_t^{\text{out}},\ \forall {t\geq 1}
\end{align}}
\end{definition}
{In this work, the outside option is typically the expected utility from selecting a server without any platform-provided information.}
Based on \Cref{def:expost_budget} and \Cref{def:exante_IR}, we propose our side-payment mechanism in the next subsection.

\subsection{Side-payment Mechanism Design and Analysis}
One can imagine that the platform's budget (thanks to charging former customers) will be used up or insufficient after rewarding some follow-up customers. Thus, we partition the entire time horizon into charge-then-reward cycles, each of which starts with charging and follows a charge-then-reward structure. Note that different charge-then-reward cycles have different periods, as they start with different system states (i.e., variable server 2's AoI $\alpha$ and queue length $k$ in the last report). 

Without loss of generality, we consider any charge-then-reward cycle $q\geq 1$ starting with reported queue length $k_q$ at server 2 to represent the entire time horizon.\footnote{{Our side-payment mechanism can be extended to the general case with $N$ independent variable servers, where the payment charged or rewarded to a customer is determined by the immediate utility difference between the server chosen under the myopic policy and the next-best alternative.}} 

\begin{definition}[Dynamic side-payment mechanism]\label{def:CFRN}
If myopic customers under-explore variable server $2$ with $A^{(m)}(k_q)>A^*(k_q)$ in the $q$-th cycle, then our dynamic side-payment mechanism operates as follows to include $A^*(k_q)$ customer arrivals in this cycle:
\begin{itemize}
    \item For customer $z$, where $1\leq z\leq A^*(k_q)-1$, the platform charges from it payment $p_z=r_1-g(z,k_q)$, without changing its choice of server~1.
    \item For customer $A^*(k_q)$, the platform gives reward ${-p_{A^*(k_q)}=g\big(A^*(k_q),k_q\big)-r_1}$ to it. This changes its server choice from server 1 to server $2$ as in the socially optimal policy. Then the new $(q+1)$-th charge-then-reward cycle starts from the next arrival.
\end{itemize}
If myopic customers over-explore variable servers with $A^{(m)}(k_q)<A^*(k_q)$ in the $q$-th cycle, then our dynamic side-payment mechanism operates below:
\begin{itemize}
    \item For customer $z$, where $1\leq z\leq A^{(m)}(k_q)-1$, the platform charges from it payment $p_z=r_1-g(z,k_q)$, without changing its choice of server 1.
    \item For customer $j$, where $j\geq A^{(m)}(k_q)$, the platform gives reward ${-p_{j}=r_1-g(j,k_q)}$ to it as long as the budget is no less than $p_j$. This changes its server choice from variable server $2$ to reliable server 1 as in the social optimum. 
    \item For the last customer $l$ in this cycle without enough budget to reward, the platform charges from it payment $p_l=g(l,k_q)-r_1$, without changing its choice of server~2. Then the new $(q+1)$-th cycle starts from the next arrival with extra budget $p_l$.
\end{itemize}
\end{definition}

According to Definition \ref{def:CFRN}, our dynamic side-payment mechanism has the following two cases to operate.

If myopic customers under-explore variable
server 2 with $A^{(m)}(k_q)>A^{*}(k_q)$ in the $q$-th charge-then-reward cycle, {the mechanism accumulates budget by charging the first $A^(k_q)-1$ customers in each cycle and only needs to reward the $A^*(k_q)$-th customer to induce exploration of server 2. As a result,}
our dynamic side-payment mechanism perfectly reduces the exploration threshold from
myopic $A^{(m)}(k_q)$ to optimal $A^{*}(k_q)$.

\begin{algorithm}[t]
\caption{{Dynamic Side-payment Mechanism}}
\label{alg:side_payment}
\begin{algorithmic}[1]
\STATE \textbf{Initialization:} Start at cycle index $q=1$, with initial state $(\alpha_q, k_q)$, and initial budget $B \leftarrow 0$
\WHILE{new customer arrives}
    \STATE Observe current state $(\alpha_q,k_q)$ and determine myopic threshold $A^{(m)}(k_q)$ and optimal threshold $A^*(k_q)$ 
    \IF{$A^{(m)}(k_q) > A^*(k_q)$} 
        \FOR{customer index $z = 1,2,\dots,A^*(k_q)-1$}
            \STATE Charge customer $z$ payment $p_z = r_1 - g(z,k_q)$
            \STATE Update budget: $B \leftarrow B + p_z$
        \ENDFOR
        \STATE Reward customer $A^*(k_q)$ payment: $-p_{A^*(k_q)} = g(A^*(k_q),k_q)-r_1$
        \STATE Update budget: $B \leftarrow B - p_{A^*(k_q)}$ 
        \STATE Reset state for next cycle $(\alpha_{q+1},k_{q+1}) \leftarrow (1, Q(k_q))$
        \STATE Increment cycle: $q \leftarrow q + 1$
    \ELSIF{$A^{(m)}(k_q) < A^*(k_q)$} 
        \FOR{customer index $z = 1,2,\dots,A^{(m)}(k_q)-1$}
            \STATE Charge customer $z$ payment $p_z = r_1 - g(z,k_q)$
            \STATE Update budget: $B \leftarrow B + p_z$
        \ENDFOR
        \STATE Set customer index $j = A^{(m)}(k_q)$
        \WHILE{$B \geq r_1 - g(j,k_q)$ \textbf{and} $j<A^*(k_q)$}
            \STATE Reward customer $j$ payment: $-p_j = r_1 - g(j,k_q)$
            \STATE Update budget: $B \leftarrow B - p_j$
            \STATE Increment customer index: $j \leftarrow j+1$
            \IF{Budget insufficient to reward customer $j$ or $j=A^*(k_q)$}
            \STATE Charge customer payment $p_j = g(j,k_q)-r_1$
            \STATE Update budget: $B \leftarrow B + p_j$
        \ENDIF
        \ENDWHILE
        \STATE Reset state for next cycle $(\alpha_{q+1},k_{q+1}) \leftarrow (1, Q(k_q))$
        \STATE Increment cycle: $q \leftarrow q + 1$
    \ELSE 
        \STATE ($A^{(m)}(k_q)=A^*(k_q)$) No payment needed, customer makes myopic/optimal choice directly
        \STATE Reset state: $(\alpha_{q+1},k_{q+1}) \leftarrow (1, Q(k_q))$
        \STATE Increment cycle: $q \leftarrow q + 1$
    \ENDIF
\ENDWHILE
\end{algorithmic}
\end{algorithm}

If myopic customers over-explore variable server 2 in the $q$-th charge-then-reward cycle, i.e., $A^{(m)}(k_q)<A^{*}(k_q)$, our mechanism increases the exploration threshold from $A^{(m)}(k_q)$ to $A^{*}(k_q)$, provided with enough budget from charging the first $A^{(m)}(k_q)-1$ customers. Yet, given a short charging period in the whole cycle (i.e., $A^{(m)}(k_q)-1<{A^*(k_q)}/{2}$), our mechanism without enough budget to reward can only increase $A^{(m)}(k_q)$ to $2 A^{(m)}(k_q)-2$ to keep ex-post budget balanced. By further charging the $(2 A^{(m)}(k_q)-1)$-th customer who explores, our mechanism can have an extra budget in the next $(q+1)$-th cycle to increase its exploration threshold by $1$. Thus, it needs $\max\left\{A^*(k_q)-\max\{2A^{(m)}(k_q)-3,1\},1\right\}$ cycles to increase the exploration threshold to the optimum.

In the next lemma, we demonstrate that our dynamic side-payment mechanism achieves ex-post budget balance for the crowdsourcing platform, ensuring its long-term sustainability, and ex-ante individual rationality for customers, thereby incentivizing their continued participation.
\begin{lemma}\label{lemma:CFRN}
{Our dynamic side-payment mechanism in \cref{def:CFRN}  satisfies the following properties:
\begin{itemize}
    \item Ex-post budget balanced at any time $t$ for the platform, that is,
\begin{align*}
    \sum_{\tau=0}^tp_{\tau}\geq 0,
\end{align*}
    where $p_{\tau}$ denotes the payment or charge associated with customer $\tau\geq t$.
    \item Ex-ante individual rationality for all customers, as defined in \eqref{EU_exante}.
\end{itemize}}
\end{lemma}

The proof of \Cref{lemma:CFRN} is given in Appendix~H of the supplemental material. 
Most crowdsourcing platforms expect customers' long-term subscriptions. 
According to \Cref{lemma:CFRN}, our budget-balanced side-payment mechanism, which both charges and rewards customers over time, results in an expected net payment of zero in the long run. 
{Consequently, the average expected net payment over all customers is zero. In addition, due to the beneficial information learning and congestion control achieved by our mechanism, customers experience a long-term increase in average utility across the population. Therefore, customers are ex-ante individually rational to participate in our platform.}
In contrast, if customers opt not to utilize the platform, they must make server choices without any information (e.g., $\alpha$ and $k$). We will further examine that this information-hiding policy results in poor average performance by experiments later in \Cref{section5-3}.

Note that the ex-post budget balance constraint does not allow our mechanism to always change myopic behaviors to achieve the social optimum (see the second case of \Cref{def:CFRN} where the myopic customers over-explore). We wonder about its efficiency ratio as compared to the social optimum. Next, we prove that our side payment mechanism can efficiently reduce $\text{PoA}^{(m)}$ in \Cref{thm:poa} and \Cref{Prop:PoA_N} to a small constant in the following.
\begin{theorem}\label{thm:incentive_poa}
Our dynamic side-payment mechanism successfully reduces the myopic policy’s infinite $\text{PoA}^{(m)}$ in (\ref{PoAm}) to $\text{PoA}^{(\$)}<2$ for any variable server number $N\geq 1$.
\end{theorem}

The proof of \Cref{thm:incentive_poa} is given in Appendix~I. In the general case with $N$ variable servers, if customers over-explore in a cycle, our mechanism will increase the AoI of all variable servers by rewarding a customer for changing its server choice to reliable server 1. Hence, our mechanism efficiently controls the expected queue length of each variable server, and thus well bounds PoA to less than $2$. 
{Note that the log-concavity of the service time distribution $F$ ensures that the incremental incentive payments required by the mechanism remain well-behaved (i.e., do not increase too rapidly) as the system state evolves. This property guarantees that the budget accumulated from charging early customers is always sufficient to reward later customers for exploration.}


Finally, one may wonder if there is a better side-payment mechanism to further reduce the bounded $\text{PoA}^{(\$)}$. We expect that it is unlikely to further reduce $\text{PoA}^{(\$)}$ under other budget-balanced side-payment mechanisms. For example, in the worst case of maximum over-exploration with $r_1\rightarrow 0$, our side-payment mechanism has charged the maximum payment $g(l,k_q)$ from customers, and thus reward the maximum number of customers to change their myopic decisions to follow the socially optimal policy.

\section{Experimental Verification Using Real-World Datasets}\label{section5-3}

In addition to the theoretical analysis of the worst-case PoA, in this section, we further conduct experiments using real-world datasets to verify our mechanism's good average performance. 

\begin{figure}[t]
    \centering
    \includegraphics[width=0.35\textwidth]{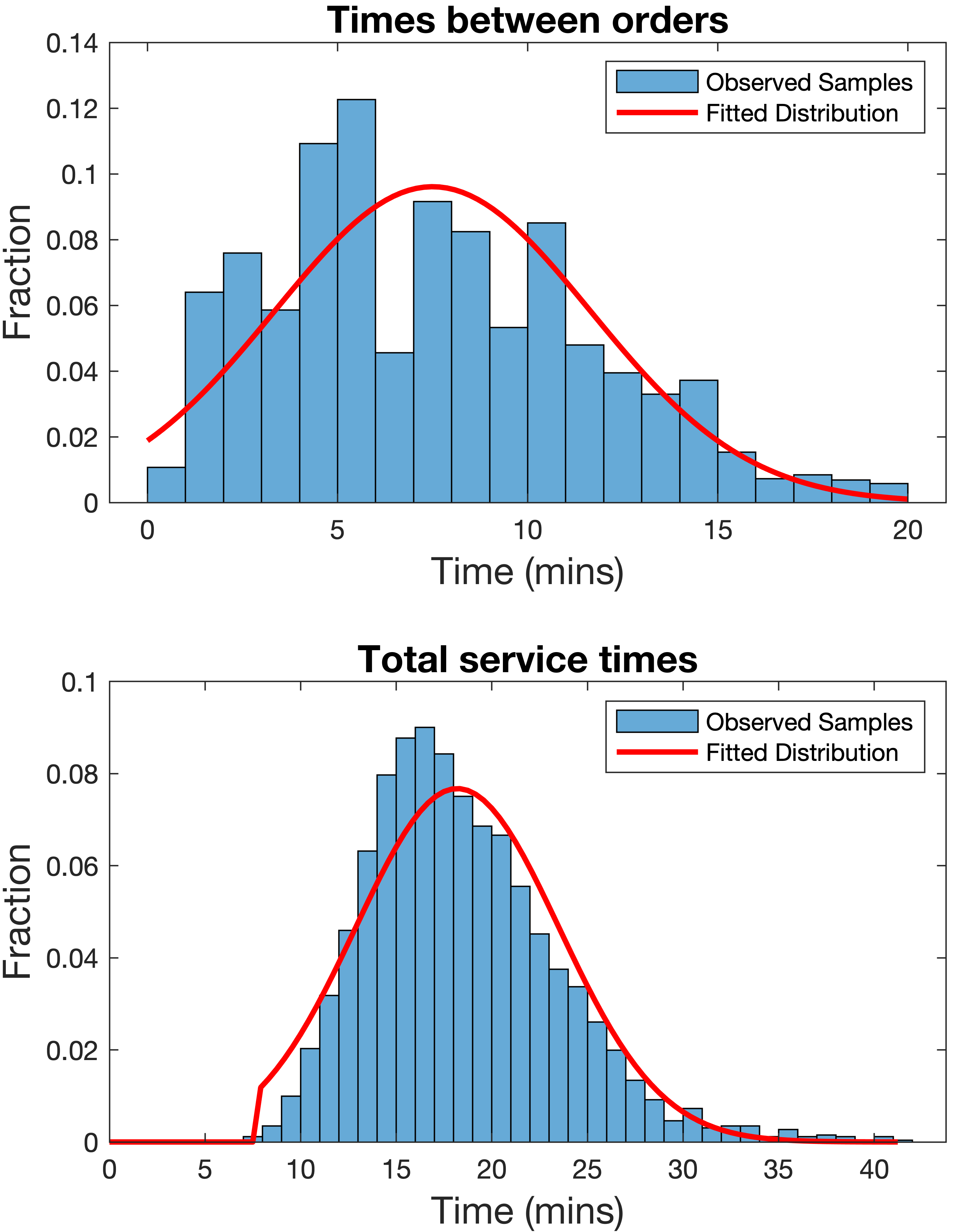}
    \caption{Data samples and the corresponding fitted normal distributions of ``times between orders" and ``total service times" in the dataset from \cite{kaggle_data}.}
    \label{fig:data_fit}
\end{figure}

First, we use real datasets from Talabat \cite{kaggle_data} to develop a practical queueing simulation for food delivery services. Specifically, we use real data to extend customers' fixed arrival intervals to random times, which follow a general distribution $G$ with an average arrival rate $\lambda$. Customers seek food delivery services during dinnertime (17:00-20:00). They have the option to cook at home without queueing, which offers a fixed benefit $r_1$. Alternatively, they may opt for food delivery from nearby restaurants. The delivery service of these restaurants has a capacity limit of $K+1$ orders, and their total service time, including the waiting time to collect food and the travel time, is random and follows distribution $F$. 

In \Cref{fig:data_fit}, we depict the data samples for ``times between orders" and ``total service times" from the dataset \cite{kaggle_data}. After analyzing the data samples, we fit their corresponding distributions $G$ and $F$ into (truncated) normal distributions (\!\!\cite{meisling1958discrete,hassin2016rational}), using maximum likelihood estimation. We obtain that customers' mean arrival time for requesting food services is $7.50$ mins with a standard deviation of $3.45$ mins. Additionally, the average total service time in restaurants is $18.52$ mins with a standard deviation of $5.01$ mins.  To align with our discrete time horizon, each slot lasting 5 minutes, we translate the mean times into rates: $\lambda=0.67$ and $\mu=0.27$, with variances $\sigma_{\lambda}^2=0.69$ and $\sigma_{\mu}^2=1.00$, respectively. Then the two distributions $G$ and $\lambda$ will be used in the following experiments. Note that our prior $F(\mu)$ can be any log-concave distribution including normal distribution to apply our analysis and mechanisms.

To compare the average performance of different policies/mechanisms, we define 
\begin{align*}
    \text{IR}^{(m)}=\frac{\mathbb{E}[V^*(\alpha,k)]}{\mathbb{E}[V^{(m)}(\alpha,k)]}
\end{align*}
to be the average inefficiency ratio of the myopic policy, where {$\mathbb{E}[V^*(\alpha, k)]$ is the expected long-term value function under the socially optimal policy and $\mathbb{E}[V^{(m)}(\alpha, k)]$ is that under the myopic policy.}
Similarly, we define the average inefficiency ratios of the popular information-hiding mechanism (\!\!\cite{li2019recommending,farhadi2022dynamic,li2025analyze}) and our dynamic side-payment mechanism to be
\begin{align*}
    \text{IR}^{(\emptyset)}=\frac{\mathbb{E}[V^*(\alpha,k)]}{\mathbb{E}[V^{(\emptyset)}(\alpha,k)]},\ \  \text{IR}^{(\$)}=\frac{\mathbb{E}[V^*(\alpha,k)]}{\mathbb{E}[V^{(\$)}(\alpha,k)]}.
\end{align*}
{Here, under the information-hiding mechanism, each customer selects servers according to a fixed probability distribution: each customer chooses server 1 with probability $0.2$, and chooses among the $N$ variable servers with equal probability, that is, each variable server is selected with probability $\frac{0.8}{N}$.}

According to \Cref{def:information_mechanism}, under the information-hiding mechanism, customers who lack knowledge of $\alpha$ and $k$ must estimate the expected utilities of variable servers based on steady-state queue lengths to make decisions.

\begin{figure}[t]
    \centering
    \includegraphics[width=0.4\textwidth]{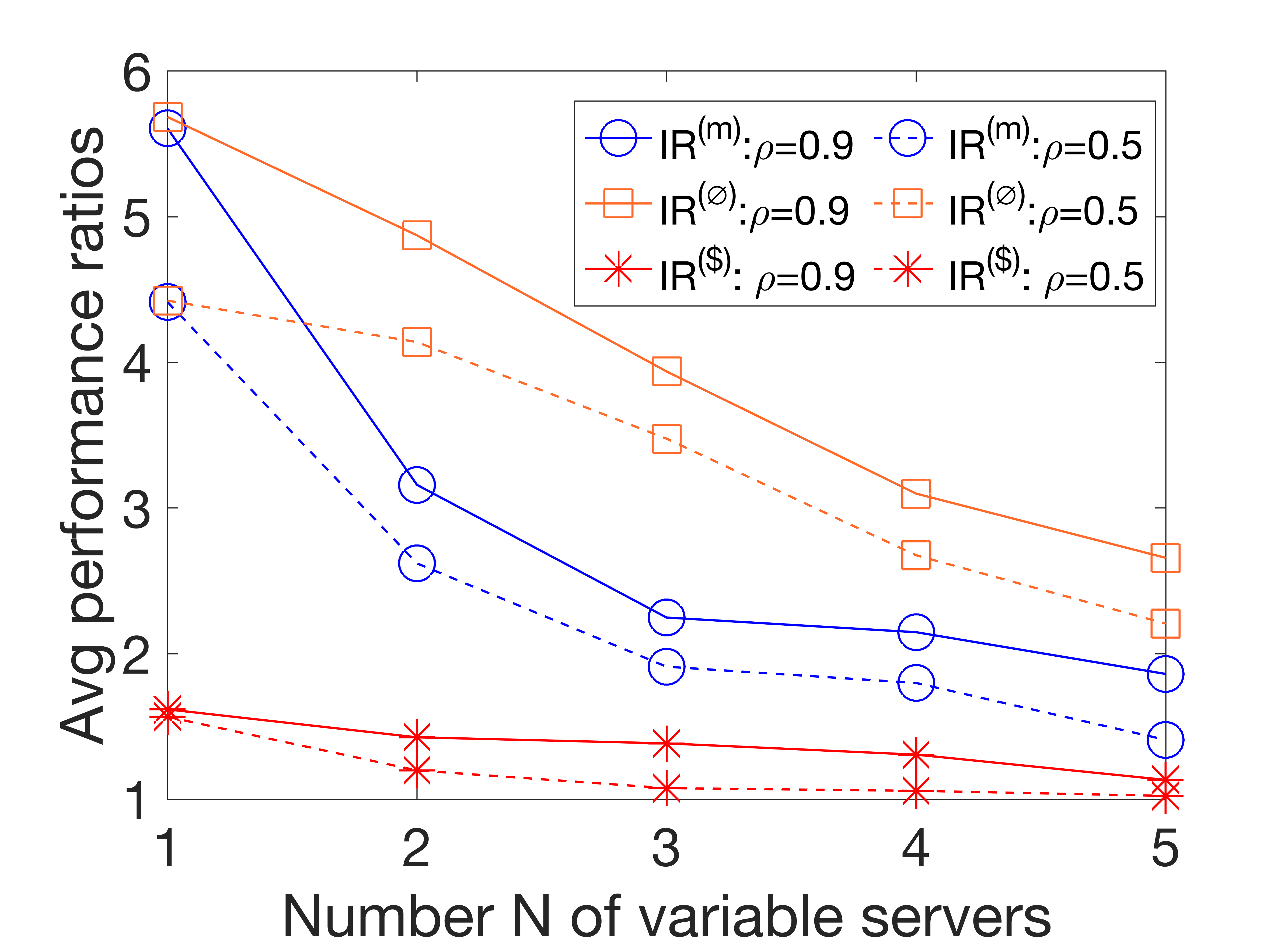}
    \caption{Average inefficiency ratios under the myopic policy, information-hiding mechanism and our side-payment mechanism plotted against server number $N\in\{1,2,3,4,5\}$ and discount factor $\rho\in\{0.9,0.5\}$. Here we set a fixed buffer size of $K=2$.}
    \label{fig:varying_rho}
\end{figure}

\begin{figure}[t]
    \centering
    \includegraphics[width=0.4\textwidth]{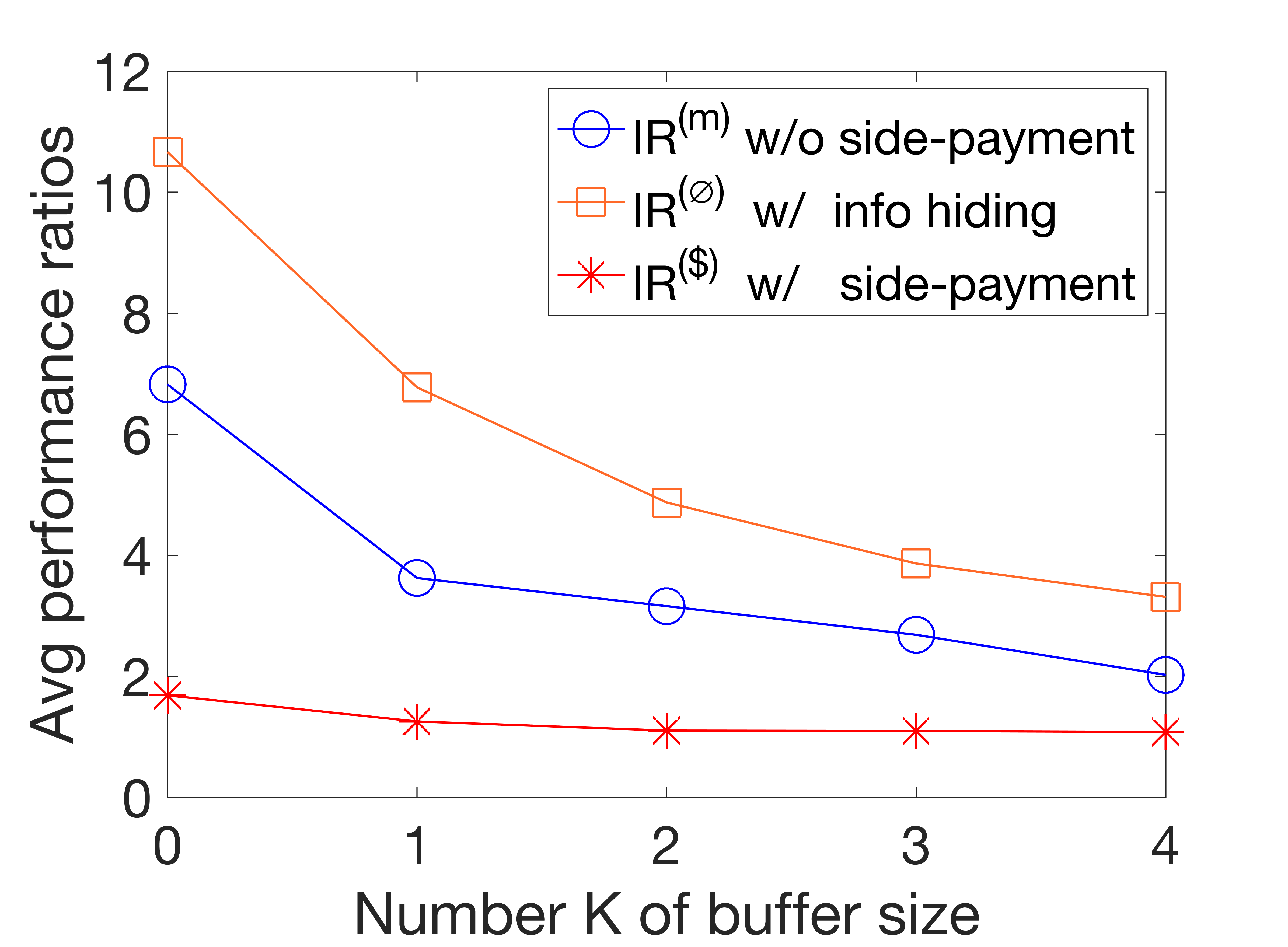}
    \caption{Average inefficiency ratios plotted against buffer size $K\in\{0, 1,2,3,4\}$, with a fixed server number of $N=2$.}
    \label{fig:varying_K}
\end{figure}

Given $\lambda=0.67, \sigma_{\lambda}^2=0.69, \mu=0.27,$ and $\sigma_{\mu}^2=1.00$ determined above, we further set $r_1=5,r_i=200,c=70,h=1$ and finite time horizon $T=50$. Initially, we define $\alpha_i(0)=1$ and $k_i(0)=K+1$ for any variable server~$i\in\{2,\cdots,N+1\}$.
After repeating each experiment $100$ times with different randomized arrival and service times, we plot Figures~\ref{fig:varying_rho} and \ref{fig:varying_K} to compare $\text{IR}^{(m)}$ and $\text{IR}^{(\emptyset)}$ to our $\text{IR}^{(\$)}$ versus variable server number $N$, discount factor $\rho$ and buffer size $K$, respectively. Based on \Cref{Prop:PoA_N}, we set all parameters, such as small $K$ and $N$, to examine particularly challenging scenarios. This ensures robust verification of our mechanism's performance.


In the first experiment, we fix the buffer size to $K=2$ for each variable server. Figure \ref{fig:varying_rho} tells that both the myopic policy and the information-hiding mechanism cause $\text{IR}^{(m)}\approx \text{IR}^{(\emptyset)}>4.4$ at the minimum $N=1$. While our dynamic side-payment mechanism successfully reduces them to $\text{IR}^{(\$)}<2$, no matter $\rho=0.9$ or $\rho=0.5$. This aligns with \Cref{thm:incentive_poa}. 
When $N=1$, myopic customers with or without the hiding mechanism always choose variable server~2, resulting in identical inefficiency ratios. While for $N\geq 2$, $\text{IR}^{(\emptyset)}$ without information sharing leads to the worst system performance, confirming our results in \Cref{Lemma:Information}.
Furthermore, all of $\text{IR}^{(m)},\text{IR}^{(\emptyset)}$ and $\text{IR}^{(\$)}$ reduce as server number $N$ or discount factor $\rho$ increases, which aligns with \Cref{Prop:PoA_N}.

In the second experiment, we further examine the average-case performance of the three policies concerning buffer size~$K$, with a fixed variable server number of $N=2$. \Cref{fig:varying_K} illustrates that the information-hiding mechanism leads to $\text{IR}^{(\emptyset)}>10$ without a buffer. This occurs because customers lacking information sharing continuously explore the two variable servers, resulting in maximum congestion there. Although the myopic policy with information sharing exhibits improved performance, $\text{IR}^{(m)}$ remains greater than $6$ when $K=0$. While our side-payment mechanism consistently maintains $\text{IR}^{(\$)}$ close to $1$. Furthermore, all values of $\text{IR}^{(m)},\text{IR}^{(\emptyset)}$ and $\text{IR}^{(\$)}$ decrease with buffer size $K$, consistent with both \Cref{thm:poa} and \Cref{Prop:PoA_N}.

\section{Conclusions}\label{section6}
In this paper, we study how to analyze and regulate human-in-the-loop learning among selfish customers in queueing systems, where the stochastic congestion information varies endogenously with customers' server choices over time. 
Our analytical result of infinite price of anarchy (PoA) implies that if customers make myopic choices on which server to select, it can result in an arbitrarily large efficiency loss. This occurs because they overexplore the server even though it may be congested. 
{Specifically, we prove that the lower bound of PoA decreases as the buffer size increases in the single-server case, while the upper bound of PoA decreases as the number of servers increases in the multi-server setting.} As the server choices of customers internally alter the queueing status, we prove that prior informational (non-monetary) mechanisms on exploration-exploitation of exogenous information make the PoA infinite. 
Hence, we propose a dynamic side-payment mechanism, which periodically charges some customers and rewards others to curb their over-exploration over time while keeping ex-post budget balanced. 
Our mechanism balances the congestion and information learning at variable servers such that PoA is reduced to less than $2$. 
In addition to the worst-case PoA analysis, we conduct experiments using real datasets to further verify that our mechanism results in good average performance.

{In future research, we aim to generalize human-in-the-loop learning (HILL) in queueing systems to the setting where the service rates of variable servers are unknown and must be learned. In such scenarios, selfish customers may be reluctant to explore servers that could have low service rates, which can result in congestion at other servers. Therefore, it is crucial to develop efficient mechanisms to regulate customers’ exploration and exploitation across different servers.
Additionally, analyzing queueing systems with heterogeneous variable servers and designing corresponding mechanisms is important to further broaden the applicability of our HILL framework.}


\begin{thebibliography}{9}

\bibitem{ouyang2012asymptotically}
W.~Ouyang, A.~Eryilmaz, and N.~B. Shroff, ``Asymptotically optimal downlink
  scheduling over markovian fading channels,'' in \emph{2012 Proceedings IEEE
  INFOCOM}.\hskip 1em plus 0.5em minus 0.4em\relax IEEE, 2012, pp. 1224--1232.

\bibitem{hassin2020queue}
R.~Hassin and R.~Roet-Green, ``On queue-length information when customers
  travel to a queue,'' \emph{Manufacturing \& Service Operations Management},
  2020.

\bibitem{wang2020efficient}
J.~Wang and M.~Hu, ``Efficient inaccuracy: User-generated information sharing
  in a queue,'' \emph{Management Science}, vol.~66, no.~10, pp. 4648--4666,
  2020.

\bibitem{chen2022food}
M.~Chen, M.~Hu, and J.~Wang, ``Food delivery service and restaurant: Friend or
  foe?'' \emph{Management Science}, 2022.

\bibitem{queue-dodging}
D.~Phelan, ``There is now an app to help you skip restaurant queues,''
  \url{https://www.timeout.com/london/blog/theres-now-an-app-to-help-you-skip-restaurant-queues-012717},
  2017.

\bibitem{Disneyland}
AppPicker, ``Wait times for {Disneyland} app review: View user-submitted
  waiting times for {Disneyland} 2021,''
  \url{https://www.apppicker.com/reviews/10509}, 2021.

\bibitem{pavemint}
Pavemint, ``Transforming the holiday park experience with smart parking,''
  \url{https://www.pavemint.com/blog/holiday-park-smart-parking}, 2022.

\bibitem{zhang2018distributed}
J.~Zhang, P.~Lu, Z.~Li, and J.~Gan, ``Distributed trip selection game for
  {Public} {Bike} {System} with crowdsourcing,'' in \emph{IEEE INFOCOM
  2018-IEEE Conference on Computer Communications}.\hskip 1em plus 0.5em minus
  0.4em\relax IEEE, 2018, pp. 2717--2725.

\bibitem{waze}
N.~Epson, ``Waze: {App} that combines {Crowdsourcing} with {GPS} navigation,''
  \url{https://www.bluelabellabs.com/blog/waze-an-app-that-combines-crowdsourcing-with-gps-navigation},
  2021.

\bibitem{myTSA}
MyTSA, ``{MyTSA Mobile Application},''
  \url{https://www.dhs.gov/publication/dhstsapia-028-mytsa-mobile-application},
  2017.

\bibitem{slivkins2019introduction}
A.~Slivkins, ``Introduction to multi-armed bandits,'' \emph{Foundations and
  Trends{\textregistered} in Machine Learning}, vol.~12, no. 1-2, pp. 1--286,
  2019.

\bibitem{liu2023nonstationary}
Y.~Liu, B.~Van~Roy, and K.~Xu, ``Nonstationary bandit learning via predictive
  sampling,'' in \emph{International Conference on Artificial Intelligence and
  Statistics}.\hskip 1em plus 0.5em minus 0.4em\relax PMLR, 2023, pp.
  6215--6244.

  
\bibitem{wang2020restless}
S.~Wang, L.~Huang, and J.~Lui, ``Restless-ucb, an efficient and low-complexity
  algorithm for online restless bandits,'' \emph{Advances in Neural Information
  Processing Systems}, vol.~33, pp. 11\,878--11\,889, 2020.

\bibitem{krishnasamy2021learning}
S.~Krishnasamy, R.~Sen, R.~Johari, and S.~Shakkottai, ``Learning unknown
  service rates in queues: A multiarmed bandit approach,'' \emph{Operations
  Research}, vol.~69, no.~1, pp. 315--330, 2021.

\bibitem{li2023congestion}
H.~Li and L.~Duan, ``When congestion games meet mobile crowdsourcing: Selective
  information disclosure,'' in \emph{Proceedings of the AAAI Conference on
  Artificial Intelligence}, vol.~37, no.~5, 2023, pp. 5739--5746.

\bibitem{li2020multi}
F.~Li, D.~Yu, H.~Yang, J.~Yu, K.~Holger, and X.~Cheng, ``Multi-armed-bandit-based spectrum scheduling algorithms in wireless networks: A survey,'' \emph{IEEE Wireless Communications}, vol.~27, no.~1, pp. 24-30, 2020.

\bibitem{li2025competitive}
H.~Li, and L.~Duan, ``Competitive Multi-armed Bandit Games for Resource Sharing,'' \emph{IEEE Transactions on Mobile Computing}, to appear, 2025.

\bibitem{li2024distributed}
H.~Li, and L.~Duan, ``Distributed learning for dynamic congestion games,'' \emph{IEEE International Symposium on Information Theory (ISIT)}, pp. 3654--3659, 2024.


\bibitem{stidham1985optimal}
S.~Stidham, ``Optimal control of admission to a queueing system,'' \emph{IEEE
  Transactions on Automatic Control}, vol.~30, no.~8, pp. 705--713, 1985.

\bibitem{legros2018m}
B.~Legros, ``M/g/1 queue with event-dependent arrival rates,'' \emph{Queueing
  Systems}, vol.~89, pp. 269--301, 2018.

\bibitem{li2019recommending}
Y.~Li, C.~Courcoubetis, and L.~Duan, ``Recommending paths: Follow or not
  follow?'' in \emph{IEEE INFOCOM 2019-IEEE Conference on Computer
  Communications}.\hskip 1em plus 0.5em minus 0.4em\relax IEEE, 2019, pp.
  928--936.

\bibitem{gaitonde2021virtues}
J.~Gaitonde, and E.~Tardos, ``Virtues of patience in strategic queuing systems'' in \emph{Proceedings of the 22nd ACM Conference on Economics and Computation}.\hskip 1em plus 0.5em minus 0.4em\relax IEEE, 2021, pp.
  520--540.

\bibitem{ibrahim2018sharing}
R.~Ibrahim, ``Sharing delay information in service systems: a literature survey,'' in \emph{Queueing Systems}.\hskip 1em plus 0.5em
  minus 0.4em\relax Springer, vol.~89, no.~1, 2018, pp. 49--79.

\bibitem{farhadi2022dynamic}
F.~Farhadi and D.~Teneketzis, ``Dynamic information design: a simple problem on
  optimal sequential information disclosure,'' \emph{Dynamic Games and
  Applications}, vol.~12, no.~2, pp. 443--484, 2022.

\bibitem{mansour2022bayesian}
Y.~Mansour, A.~Slivkins, V.~Syrgkanis, and Z.~S. Wu, ``Bayesian exploration:
  Incentivizing exploration in bayesian games,'' \emph{Operations Research},
  vol.~70, no.~2, pp. 1105--1127, 2022.

\bibitem{guo2022signaling}
P.~Guo, M.~Haviv, Z.~Luo, and Y.~Wang, ``Signaling service quality through
  queue disclosure,'' \emph{Manufacturing \& Service Operations Management},
  2022.

\bibitem{borgers2015introduction}
T.~B{\"o}rgers and D.~Krahmer, \emph{An introduction to the theory of mechanism
  design}.\hskip 1em plus 0.5em minus 0.4em\relax Oxford University Press, USA,
  2015.

\bibitem{li2017dynamic}
Y.~Li, C.~A. Courcoubetis, and L.~Duan, ``Dynamic routing for social
  information sharing,'' \emph{IEEE Journal on Selected Areas in
  Communications}, vol.~35, no.~3, pp. 571--585, 2017.

\bibitem{ferguson2021effectiveness}
B.~L. Ferguson, P.~N. Brown, and J.~R. Marden, ``The effectiveness of subsidies
  and tolls in congestion games,'' \emph{IEEE Transactions on Automatic
  Control}, 2021.

\bibitem{ma2021spatio}
H.~Ma, F.~Fang, and D.~C. Parkes, ``Spatio-temporal pricing for ridesharing
  platforms,'' \emph{Operations Research}, 2021.

\bibitem{tavafoghi2017informational}
H.~Tavafoghi and D.~Teneketzis, ``Informational incentives for congestion
  games,'' in \emph{2017 55th Annual Allerton Conference on Communication,
  Control, and Computing (Allerton)}.\hskip 1em plus 0.5em minus 0.4em\relax
  IEEE, 2017, pp. 1285--1292.

\bibitem{zhang2020traffic}
Q.~Zhang, S.~Q. Liu, and M.~Masoud, ``A traffic congestion analysis by user
  equilibrium and system optimum with incomplete information,'' \emph{Journal
  of Combinatorial Optimization}, pp. 1--14, 2020.

\bibitem{lin1984optimal}
W.~Lin and P.~Kumar, ``Optimal control of a queueing system with two
  heterogeneous servers,'' \emph{IEEE Transactions on Automatic control},
  vol.~29, no.~8, pp. 696--703, 1984.

\bibitem{meisling1958discrete}
T.~Meisling, ``Discrete-time queuing theory,'' \emph{Operations Research},
  vol.~6, no.~1, pp. 96--105, 1958.

\bibitem{hassin2016rational}
R.~Hassin, \emph{Rational queueing}.\hskip 1em plus 0.5em minus 0.4em\relax CRC
  press, 2016.

\bibitem{bedewy2019minimizing}
A.~Bedewy, Y.~Sun and N.~B. Shroff, ``Minimizing the age of information through queues,'' \emph{IEEE Transactions on Information Theory}, vol.~65, no.~8, pp. 5215--5232,
  2019.
  
\bibitem{li2022online}
H.~Li and L.~Duan, ``Online pricing incentive to sample fresh information,'' \emph{IEEE Transactions on Network Science and Engineering}, vol.~10, no.~1, pp. 514--526,
  2022.

\bibitem{lippman1975applying}
S.~A. Lippman, ``Applying a new device in the optimization of exponential
  queuing systems,'' \emph{Operations Research}, vol.~23, no.~4, pp. 687--710,
  1975.

\bibitem{bellman1966dynamic}
R.~Bellman, ``Dynamic programming,'' \emph{Science}, vol. 153, no. 3731, pp.
  34--37, 1966.

\bibitem{powell2007approximate}
W.~B.~Powell, ``Approximate Dynamic Programming: Solving the curses of dimensionality,'' \emph{John Wiley \& Sons}, vol. 703, 2007.


\bibitem{sutton1998reinforcement}
R.~Sutton, and A.~G.~Barto, ``Reinforcement learning: An introduction,'' \emph{MIT press Cambridge}, vol. 1, no. 1, 1998.

\bibitem{koutsoupias1999worst}
E.~Koutsoupias and C.~Papadimitriou, ``Worst-case equilibria,'' in \emph{Annual
  symposium on theoretical aspects of computer science}.\hskip 1em plus 0.5em
  minus 0.4em\relax Springer, 1999, pp. 404--413.

\bibitem{kosenok2008individually}
G.~Kosenok and S.~Severinov, ``Individually rational, budget-balanced
  mechanisms and allocation of surplus,'' \emph{Journal of Economic Theory},
  vol. 140, no.~1, pp. 126--161, 2008.

\bibitem{alva2020impossibility}
S.~Alva and V.~Manjunath, ``The impossibility of strategy-proof, pareto
  efficient, and individually rational rules for fractional matching,''
  \emph{Games and Economic Behavior}, vol. 119, pp. 15--29, 2020.

\bibitem{kaggle_data}
Talabat, ``{Talabat Dubai detailed data for Feb 2023},''
  \url{https://www.kaggle.com/datasets/zivaank/data-feb-2023?resource=download},
  2023.

\bibitem{li2025analyze}
H.~Li, and L.~Duan, ``To Analyze and Regulate Human-in-the-Loop Learning for Congestion Games,'' \emph{IEEE Transactions on Networking}, to appear, 2025.

\end{thebibliography}





\begin{IEEEbiography}[{\includegraphics[width=1in,height=1.25in,clip,keepaspectratio]{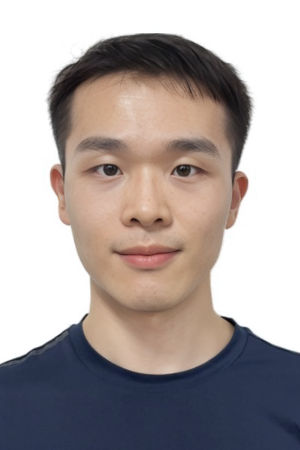}}]{Hongbo Li}(M'24)
received the B.Sc. degree from Shanghai Jiao Tong University (SJTU), China, in 2019, and the Ph.D. degree from Singapore University of Technology and Design (SUTD), Singapore, in 2024. He is currently a Postdoctoral Scholar with the Department of Electrical and Computer Engineering (ECE), The Ohio State University (OSU), Columbus, OH, USA, where he was previously a Visiting Scholar in 2024. His research interests include machine learning, networked AI, game theory, and mechanism design.
\end{IEEEbiography}

\vspace{11pt}
\begin{IEEEbiography}[{\includegraphics[width=1in,height=1.25in,clip,keepaspectratio]{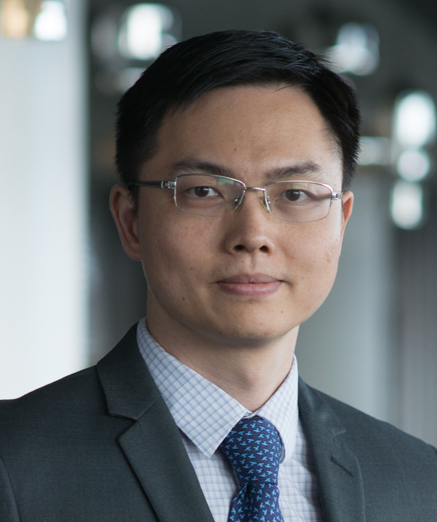}}]{Lingjie Duan}(S'09-M'12-SM'17) received the Ph.D. degree from The Chinese University of Hong Kong in 2012. He is an Associate Professor at the Singapore University of Technology and Design (SUTD) and is an Associate Head of Pillar (AHOP) of Engineering Systems and Design. In 2011, he was a Visiting Scholar at University of California at Berkeley, Berkeley, CA, USA. His research interests include network economics and game theory, network security and privacy, energy harvesting wireless communications, and mobile crowdsourcing. He is an Associate Editor of IEEE/ACM Transactions on Networking and IEEE Transactions on Mobile Computing. He was an Editor of IEEE Transactions on Wireless Communications and IEEE Communications Surveys and Tutorials. He also served as a Guest Editor of the IEEE Journal on Selected Areas in Communications Special Issue on Human-in-the-Loop Mobile Networks, as well as IEEE Wireless Communications Magazine. He served as a General Chair of WiOpt 2023 Conference and is a regular TPC member of some other top conferences (e.g., INFOCOM, MobiHoc, SECON). He received the SUTD Excellence in Research Award in 2016 and the 10th IEEE ComSoc Asia-Pacific Outstanding Young Researcher Award in 2015.
\end{IEEEbiography}

\vspace{11pt}
\begin{IEEEbiography}
[{\includegraphics[width=1in,height=1.25in,clip,keepaspectratio]{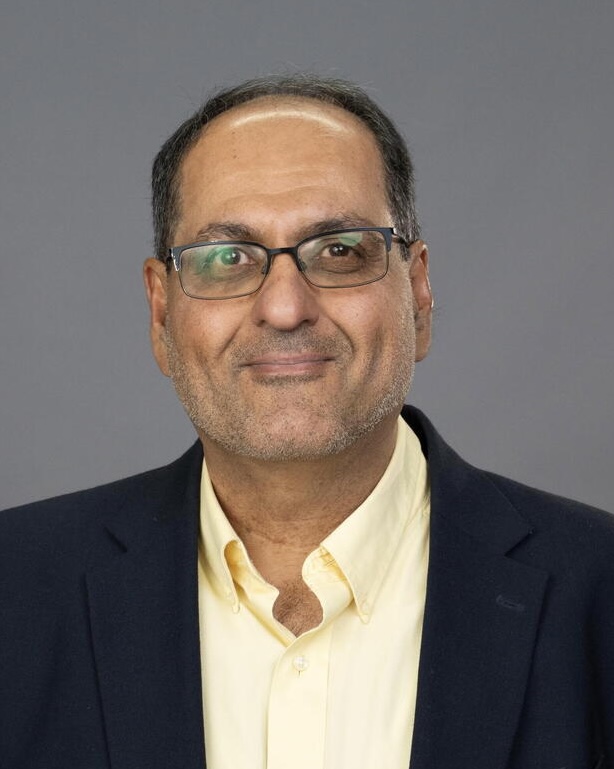}}]{Ness B. Shroff}(S'91–M'93–SM'01–F'07) received the Ph.D. degree in electrical engineering from Columbia University, New York, NY, USA, in 1994. He joined Purdue University, West Lafayette, IN, USA, immediately thereafter as an Assistant Professor with the School of Electrical and Computer Engineering. At Purdue, he became a Full Professor of ECE and the Director of a University-Wide Center on Wireless Systems and Applications in 2004. In 2007, he joined The Ohio State University, Columbus, OH, USA, where he holds the Ohio Eminent Scholar Endowed Chair in networking and communications, with the Departments of ECE and CSE. He currently serves as the Institute Director of the NSF AI Institute for Future Edge Networks and Distributed Intelligence (AI-EDGE). He is also the director of the newly formed AI$^{\text{(X)}}$ Hub at The Ohio State University. He holds or has held Visiting (chaired) Professor positions with Tsinghua University, Beijing, China, Shanghai Jiao Tong University, Shanghai, China, and the Indian Institute of Technology Bombay, Mumbai, India. He was the recipient of numerous best paper awards for his research and is listed in Thomson Reuters’ on The World’s Most Influential Scientific Minds, and has been noted as a Highly Cited Researcher by Thomson Reuters in 2014 and 2015. He served as the Editor in Chief of the IEEE/ACM Transactions on Networking, and currently serves as the Steering Committee Chair of ACM Mobihoc. He also was the recipient of the IEEE INFOCOM Achievement Award for seminal contributions to scheduling and resource allocation in wireless networks.

\end{IEEEbiography}

\vfill

\end{document}